# Improved Content Based Image Watermarking

Arvind Kumar Parthasarathy

**Abstract**

Due to improvements in imaging technologies and the ease with which digital content can be created and manipulated, there is need for the copyright protection of digital content. It is also essential to have techniques for authentication of the content as well as the owner. To this end, this paper proposes a robust and transparent scheme of watermarking that exploits the human visual systems' sensitivity to frequency, along with local image characteristics obtained from the spatial domain, improving upon the content based image watermarking scheme of Kay and Izquierdo [1]. We implement changes in this algorithm without much distortion to the image, while making it possible to extract the watermark by use of correlation. The underlying idea is generating a visual mask based on the human visual systems' perception of image content. This mask is used to embed a decimal sequence [2], [3], [4] while keeping its amplitude below the distortion sensitivity of the image pixel.

We consider texture, luminance, corner and the edge information in the image to generate a mask that makes the addition of the watermark less perceptible to the human eye. The operation of embedding and extraction of the watermark is done in the frequency domain thereby providing robustness against common frequency-based attacks including image compression and filtering. We use decimal sequences for watermarking instead of pseudo random sequences, providing us with a greater flexibility in the choice of sequence. Weighted Peak Signal to Noise Ratio is used to evaluate the perceptual change between the original and the watermarked image.

## 1 Frequency based Watermarking

To obtain better imperceptibility as well as robustness, the addition of the watermark is done in a transformed domain [5], [6], [7], [8]. DCT and DWT are two such popular transforms, operating in the frequency domain. Frequency-based techniques are very robust against attacks involving image compression and filtering because the watermark is actually spread through out the image, not just operating on an individual pixel. This is just one of the many advantages of embedding the watermark in a transformed domain as opposed to watermarking in the spatial domain. It is also well known as to how efficiently the transformed coefficients can be altered in order to minimize perceptual/visual distortion in the watermarked image which explains why such schemes are widely implemented.

In a frequency-based watermarking scheme, the watermark, upon inverse transformation to the spatial domain, is dispersed throughout the image making it very difficult for an attacker to remove the watermark without causing significant damage to the image. Watermarking in the DCT domain is usually performed on the lower or the mid-band frequencies, as higher frequencies are lost when the image is compressed. DCT watermarking can be done for an entire image taken together or block-wise. In both these methods the image is transformed into its DCT coefficients and the watermark is added to these coefficients. Finally the watermarked coefficients are inverse-transformed into the spatial domain thereby spreading the watermark throughout the image or a block of the image. In almost all the transformed domain watermarking techniques there is a trade-off between robustness and imperceptibility. The watermark, if embedded in the perceptually significant components, would result in a visible change in the final watermarked image. On the other hand, if it were embedded in the perceptually insignificant components then they would not be as robust and hence less resilient to most attacks.

## 2 Correlation-based Watermarking

Schemes are divided into blind, semi-blind and non-blind watermarking schemes based on the requirement of the original image at the receiver. In most schemes, the watermark is typically a pseudo randomly generated noise sequence, which is detected at the receiver using correlation. The generalized algorithm of most correlation-based spread spectrum watermarking in a spatial domain is based on the following equation:
$$WI (i, j) = I (i, j) + k \times W (i, j)$$
where WI = watermarked image, I = original image, k = scaling factor, and W is the watermark.

The watermark W is a pseudo randomly generated noise, based on a secret key. The main requirement for a correlation-based algorithm is that the noise should be uniformly distributed and both the noise and the image content should not be correlated. At the receiver, the correlation is done between the noise generated using the same key and the possibly altered watermarked image. If this value is greater than a pre-determined threshold then the watermark is said to be present. There is a trade off between the imperceptibility of the watermark and the scaling factor because the greater the value of the scaling factor, the higher is the probability of not making an error in the detection of the watermark. By choosing an optimum value for scaling factor, this method, although prone to errors, can to a large extent determine the presence of a watermark effectively.

Much research has been done to increase the robustness and the data hiding capacity of watermarking techniques based on perceptual properties of the Human Visual System (HVS) [1], [5], [9], [10]. The development and improvement of accurate human vision models helps in the design and growth of perceptual masks that can be used to better hide the watermark information thereby increasing its security.

Most steganographic techniques that are designed to be robust insert the watermark information into the cover image in a way that is perceptually significant. Other techniques that are relatively better at hiding information, like the LSB method, are highly vulnerable to having the embedded data distorted or quantized by lossy image compressions like JPEG. For obvious reasons, we will consider an invisible watermarking method that is capable of hiding the watermark information in the cover image in an unnoticeable way. This imperceptibility is obtained by considering the various properties of the HVS that make the scheme more robust to many types of attacks. Existing algorithms for watermarking still images usually work either in spatial domain or in transformed domain.

Our watermarking scheme deals with the extraction of the watermark information in the absence of the original image. Hence we make use of correlation-based watermark detection. A decimal sequence is added, to the cover object, instead of a PN sequence, based on the actual watermark. The results and formulae are based on a $512 \times 512$ size cover image and a block refers to a DCT block of size $8 \times 8$, which is used for better robustness against JPEG compression.

## 3 Proposed Scheme

Our scheme utilizes the perceptual information of the image content, by taking advantage of frequency selectivity and assigns weights to provide some perceptual criteria in the watermarking process. This directly results in providing the watermark more invisibility. DCT based watermarking is resistant to compression and other frequency-based attacks, and this results in the scheme being very robust as well as imperceptible than most other schemes. We divide our scheme into three steps namely: 1) generation of a mask based on the perceptual properties of the image; 2) watermarking, by spreading the d-sequence in the frequency domain, by multiplying it with the weights calculated from step1; and 3) extraction of the watermark by using a correlation-based method.

### 3.1 Just Noticeable Distortion (JND) Visual Mask

Just Noticeable Distortion is defined as a measure referring to the capability of a human observer to detect noise or distortion in the field of view. The image is first analyzed both in the frequency domain as well as spatial domain to detect the distortion sensitivity of the image based on its content. Most schemes regard the process of watermarking as adding noise to an image. An image can be distorted only to a certain limit without making the difference between the original image and the watermarked one perceptible [1]. The limit to which we can alter a pixel value without making it perceptible is the JND. There are many characteristics that define the JND, of which we consider, texture, luminance, edge and corners to estimate a mask, which is the weight assigned to the particular block. This weight is used to modulate the watermark thereby keeping the amplitude of the signal below the noise distortion sensitivity of each pixel.

Our model takes into account an image independent part based on frequency sensitivity and an image dependent part based on edge and corner information, and the luminance sensitivity [11] in designing the perceptual mask. We first segment the image into blocks based on the frequency characteristics, as the human eye is sensitive to certain frequencies more than the others. In other words when we perform the DCT on the image the resultant will be DCT coefficients arranged in a specific order based on the frequency. The image characteristics that are considered to generate the mask are texture, edge, corner and luminance. Several studies on the HVS have shown that in highly textured areas the distortion visibility is low. These areas are suited to hide the watermark signal and therefore the JND values corresponding to those areas must be high. Edge, corner and the luminance sensitivity values that are generated from the spatial domain are considered as equally important characteristics that influence the human perception of images and they have the lowest JND values [1].

In theory, the definition of a good JND mask would depend on the accurate extraction of the luminance, texture, edge and corner information from the image as this will provide maximum strength/robustness, high capacity and imperceptibility. Our scheme is image adaptive as it incorporates the local information extracted from the image. This provides adaptability, by allowing more watermark information to be embedded in blocks that have high texture. Image blocks having many edges or corners are assigned lower JND values because in these blocks the watermark can be more easily perceived. The algorithm that is used to extract the DCT coefficients is explained below:

Let $f(x,y)$ be the original grey scale cover image. This image is segmented into non-overlapping blocks of size $8 \times 8$. This is denoted as $B_k, n = 0,1,2,...N-1$.

$$f(x,y) = \bigcup_{n=0}^{N-1} B_n = \bigcup_{n=0}^{N-1} f_n(i,j), \text{ where } 0 \leq i,j < 8$$

The formula to calculate a two dimensional DCT in MATLAB is given by:

$$B_{pq} = \alpha_p \alpha_q \sum_{m=0}^{M-1} \sum_{n=0}^{N-1} A_{mn} \cos\frac{\pi(2m+1)p}{2M} \cos\frac{\pi(2n+1)q}{2N}, \begin{array}{l} 0 \leq p \leq M-1 \\ 0 \leq q \leq N-1 \end{array}$$

$$\alpha_p = \begin{cases} 1/\sqrt{M} & p=0 \\ \sqrt{2/M} & 1 \leq p \leq M-1 \end{cases} \qquad \alpha_q = \begin{cases} 1/\sqrt{N} & q=0 \\ \sqrt{2/N} & 1 \leq q \leq M-1 \end{cases}$$

where $M$ and $N$ are the row and the column size of A respectively.

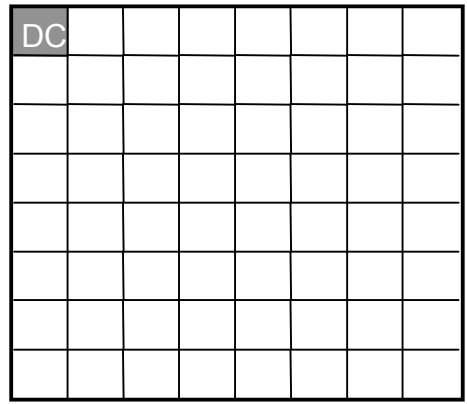

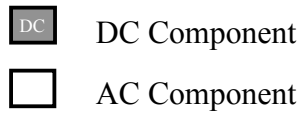  DC Component

□  AC Component

Figure 1 Definition of DC/AC components for a $8 \times 8$ DCT block

Each element of the two-dimensional array of frequency components represents a two-dimensional frequency component. The element in the upper-left corner is the DC coefficient for the entire array and all the remaining coefficients containing the frequency information are called the AC components. The DC coefficient is proportional to the average pixel value in the original block and the AC coefficients in a block describe their variation around the DC value The coefficients close to the DC component represent the highly correlated pixel values i.e., the low frequency while the coefficients towards the lower right corner represent the high frequency such as the edges and the noise.

Texture: It is defined as the visual quality of the surface of the object, exposed in an image by variances in tone, depth and shape. On obtaining the DCT coefficients, we first extract the texture information directly by analyzing these coefficients. This information is calculated from the DCT coefficients i.e., derived from the visual model consisting of an image independent part based on frequency sensitivity. Each $8 \times 8$ block which consists of the 64 DCT coefficients is analyzed and as we know that the highly textured regions or along edges the energy of the signal is concentrated in the high frequency components while in areas where the image is uniform the energy of the signal is concentrated in the low frequency components. To determine a measure for the texture information within each block based on the energy in the AC coefficients we use the formula:

$$P_T = \log(\sum_{i=1}^{63} v_i^2 - v_0^2)$$

where $v_i, i = 0, 1\ldots63$ are the 64 DCT coefficients of the $8 \times 8$ block that is being considered. We must note that $v_0$ is the value of the DC component of the DCT coefficients and it is not considered when calculating the texture value. For each block the obtained values of $E_T$ are first scaled to the range of [0, 64] and then the normalized values are assigned to the corresponding blocks.

$$M_T = \frac{64 \times P_T}{\max(P_T)}$$

Hence for an image matrix of size $512 \times 512$ we will have a matrix $E_T$ of size $64 \times 64$ where each one of those values corresponds to the texture information of each $8 \times 8$ block.

Edge: Edges are extracted from the pixel domain and this information is useful in determining the amount of watermark information that can be embedded in the image. We need an algorithm that accurately extracts the edge information from the image and by accurate we mean something that not only shows, as many edges as there are present but also differentiate between real edges and spurious edges, which may occur due to noise and texture. Human eye is more sensitive to changes in areas having more edges as compared to those with lesser or no edges (smooth areas) and hence we will use this information to assign a weight accordingly. In order to perform this we will make use of an algorithm implemented by Peter Kovesi [12], [13] that has been proved to extract edges better than most other edge detection algorithms.

There are many methods of finding the edges based on either the gradient of the image *I (u, v)* or the zero crossings, after filtering the image *I (u, v)* with a Gaussian or a Laplacian filter. One of the better algorithms returning many more edges with good accuracy in determining spurious edges is based on the phase congruency of feature detection [12]. This method is invariant to image contrast, unlike most methods. Phase congruency is described as a dimensionless quantity that provides the information that does not change with image contrast and this is used to determine the principal magnitudes of moments of phase congruency. An edge would be one where the maximum moment of phase congruency is large. More information on phase congruency and extraction of edges using phase congruency can be found in [13]. This algorithm can also be used to detect a corner map, which is strictly a subset of the edge map. Although corners are a part of an edge it is found that most edge detectors do not accurately detect edges at a corner and hence we will use a separate algorithm to detect the corners. We are concerned with determining as many edges and corners without one being dependent on the other and the rationale behind this, is explained later in this chapter. We will take an average of these two values to estimate the JND mask. Using a binary edge map, we calculate the normalized edge information for each block using the formula:

$$M_E = \frac{64 \times P_E}{\max(P_E)}$$

where $P_E$ is the cardinality of set of pixels at edge locations in each block while max ($P_E$) is the maximum value of $P_E$ over the entire image. Detected edges of the image Lena, by the two different methods mentioned earlier are shown in Figures 2 and 3. We clearly observe from these two images that the number of edges that are detected using phase congruency is far more than those detected by the Canny operator. It is also more accurate in distinguishing between real and spurious edges.

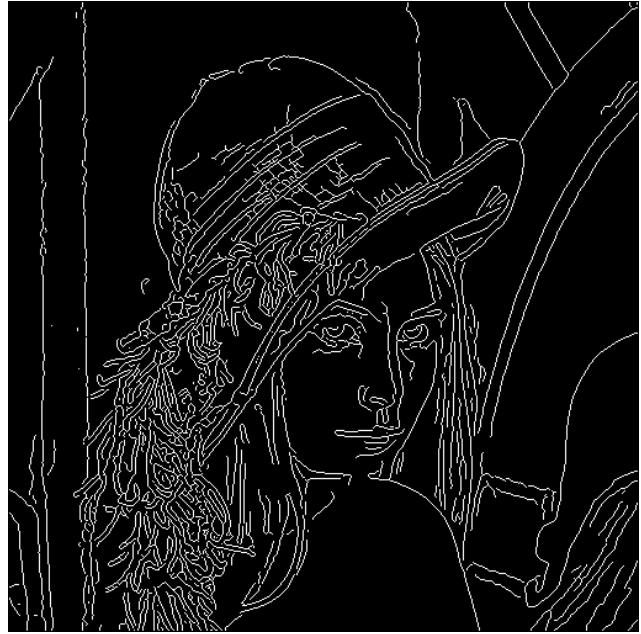

Figure 2 Edge Extraction using Canny Operator

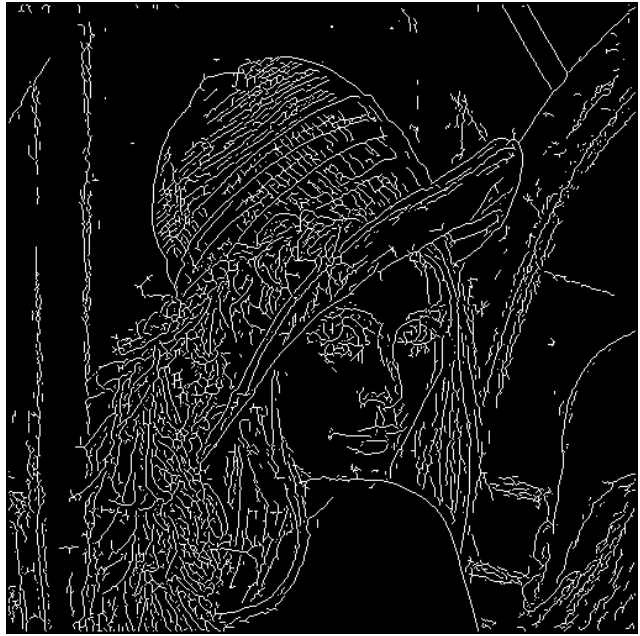

Figure 3 Edge Extraction based on Phase Congruency

Corner: Another important aspect in the pixel or spatial domain is information pertaining to the corners. Corners have long been recognized as visual information carriers and various algorithms have been proposed to detect them for use in basic visual tasks [14]. Corners are considered more localized than edges and are better in defining shapes of objects in images as an edge can provide local information only in one direction, normal to the edge [14]. A corner represents the point where two edges meet and the human eye is more sensitive to changes made in these places. Perceptual watermark schemes consider uniformity as an important factor in human perception. Kay and Izquierdo make use of a Moravec operator to extract uniform regions. It is essentially a corner detector that uses a sliding window approach to detect the smoothness in a block with the help of intensity variation. The number of pixels belonging to a uniform area in a block is regarded as the uniformity factor [1]. But a Moravec operator is found to identify false corners especially at isolated pixels, due to its sensitivity to noise [15]. The number of corners in a block accurately represents the uniformity factor and we will utilize an improved corner detector to determine this factor. There are many algorithms using different approaches to detect the right corners while eliminating the false corners and we need an algorithm that detects all the true corners that are present in an image accurately, reducing the probability of detecting false corners. The algorithm should be robust with respect to noise and the corner points should be well localized. We make use of an improved corner detection algorithm based on curvature scale space (CSS) with adaptive threshold and dynamic region of support in order to detect corners in the image [16], [17].

Methods employing CSS to detect corners have been very successful and it is believed to perform better than the existing corner detectors [15], [16]. The main steps involved in corner detection are listed below [16]:

- Extracting the edge information/contours from a binary edge map that is obtained using any good edge detection method, in our case, is by using the algorithm by Peter Kovesi.
- Filling in the gaps in the contours.
- Computing the curvature at a fixed low scale to retain all the true corners.
- Finally, the curvature local maxima are considered as corners while eliminating the rounded and false corners resulting from noise using adaptive threshold and the angle of corner.

Figure 4 shows the effective detection of corners for a Lena image based on the curvature scale space. On obtaining the corners by the above method we calculate the corner information for each block of the image using the formula:

$$M_C = \frac{64 \times P_C}{\max(P_C)}$$

where $P_C$ is the cardinality of the group of pixels determined to be corners in each block, max ($P_C$) is the maximum value of $P_C$ over all the blocks in the image and $M_C$ is the normalized value of the corner information.

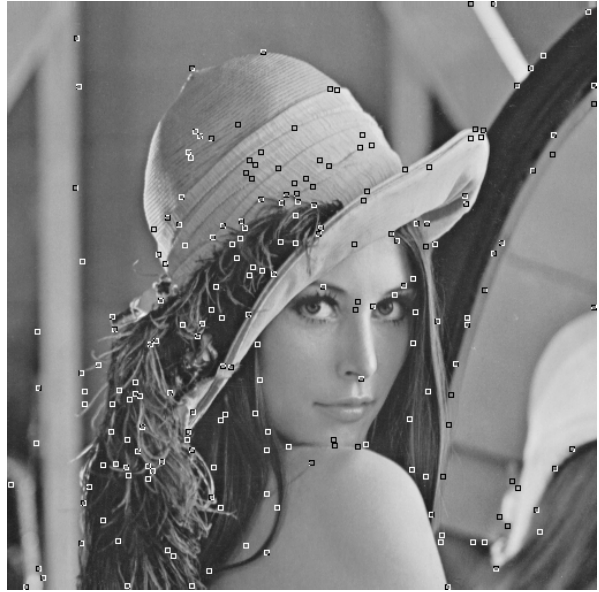

Figure 4 Corner Detection using Curvature Scale Space Method

Luminance: It is defined as the way the human eye perceives brightness of different colors. This property influences the perception of the image information by the human eye. This factor may be determined in two different domains, the frequency domain where the DC component of the DCT coefficients is used as developed by the Watson model [18], and the pixel domain. The DC component carries significant information with respect to the luminance in the block and it determines the average brightness in the block. This mean value of luminance of a local block is estimated by the formula:

$$D_L = \left\{\frac{DC_b}{DC_{mean}}\right\}^\alpha$$

where $DC_b$ is the DC coefficient of the DCT for block $b$
$DC_{mean}$ is the DC coefficient of the mean luminance of the display
$\alpha$ is the parameter that is used to control the luminance sensitivity

The value of $\alpha$ is set to 0.649 as per the model used by the authors [19]. The luminance factor is generated in both the frequency domain as per the Watson model as well as in the pixel domain and hence it is partly image dependent i.e., it is obtained by analyzing the pixels in the block. Our scheme utilizes the luminance factor that is calculated by measuring the average pixel value of the gray scale image for that block.

$$M_L = \frac{P_L}{64}$$

where $P_L$ is the sum of all the pixel values in the block and $M_L$ is the average of the luminance values within the considered block. The factors obtained from the edges and the corners and the luminance values of the image put together are called as the spatial masking values, as they are obtained directly by analyzing the pixels in the spatial domain. After obtaining the four values corresponding to the texture, edge, corners and the luminance we generate the initial mask using the equation:

$$J_I = M_T - \frac{1}{2}(M_E + M_C)$$

The human vision system is more sensitive to the changes in intensity in the mid-gray region and it is to be noted that this sensitivity fails parabolically at both ends of the gray scale. Hence a correction to the initial JND parameter value is introduced and the final JND parameter value for each block is calculated as:

$$J_F = J_I + (128 - M_L)^2$$

where $J_I$ is the initial JND parameter value, $M_L$ is the average of the luminance values within the considered block. An alternate method is to multiply the luminance factor $D_L$, derived in the frequency domain, with the JND value $J_I$ generated above and the DCT coefficient, at the time of watermark embedding. Effectively the JND value for each block is obtained by analyzing properties like texture from the frequency domain and some basic properties derived from the spatial domain namely, edge, corner and the average luminance value.

### 3.2 Watermark Embedding

We then perform the watermark insertion in DCT domain by modifying selected DCT coefficients, which embeds a d-sequence based on the watermark, for each block. The JND value controls the strength of watermark for each block. In other words, the strength of the watermark component embedded, in a block with a low JND value, is less. This is because any changes made in this block are more perceptible to the human eye. On the other hand, the strength of the watermark component embedded in a block with a high JND value is low.

The whole idea of block based JND watermarking is to incorporate the local perceptual masking effects as it provides local control of the strength of watermark based on the image content. Experimental results have shown that embedding the watermark in the high frequency components that carry less perceptually significant information results in the removal of the watermark through compression and noise attacks, while adding it in the low frequency components, which carries perceptually important information, results in visible changes in the watermarked image. In our scheme we will select and modify only those DCT coefficients that lie in the mid-frequency band.

The location of mid-frequency components of a $8 \times 8$ DCT block is shown below in Figure 5. The first value is called the DC component of the image and its DCT coefficient is relatively very high as compared to all the other coefficients, which are called the AC coefficients. Also the AC coefficients closer to the DC value comprise of the low frequency components while the ones at the bottom right are the higher frequency components.

The DC component of a DCT block is considered to carry perceptually significant information. It is believed that the DCT coefficients in the mid-frequency band have similar magnitudes. For each $8 \times 8$ transformed block the d-sequence multiplied by a scaling factor and the JND mask is added into the selected mid-frequency DCT components while the low and high frequency coefficients are copied over unaffected.

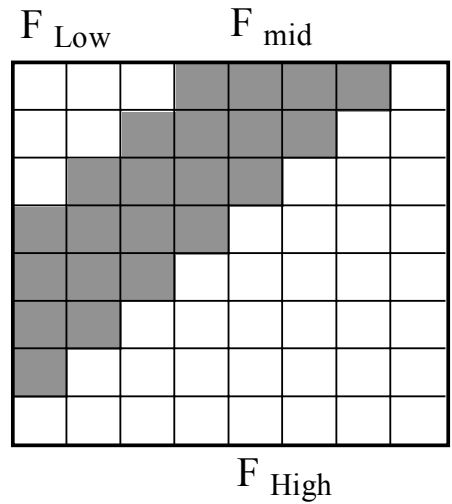

Figure 5 Definition of DCT regions

Our scheme also differs from that of Kay and Izquierdo's scheme in the way the random sequence is added to the image. The d-sequence is dependent on the actual watermark i.e., each DCT block of size $8 \times 8$ corresponds to either a 1 or 0 of the watermark bit and the d-sequence is added to the block if the watermark bit is 0 and it is subtracted wherever the watermark bit is 1. This increases both the robustness and the capacity of image to carry more watermark information while keeping the PSNR constant as compared to other spatial CDMA spread-spectrum watermarking methods where the distortion of the watermarked image increases exponentially with the size of the watermark.

The scaling factor denotes the strength of the watermark and it can be used to control the overall robustness and the quality of the image. Increasing this value increases the strength of the watermark but introduces a gradual visible distortion while decreasing this value would result in better hiding of the watermark and hence better quality but decreases the strength of the watermark. An optimal value needs to be decided upon depending on the watermark and the d-sequence before embedding. Upon inverse transformation the watermark will be scattered over the entire image and we obtain the watermarked image. The watermark embedding is done using the formula:

$$I_w(u,v,b) = \begin{cases} I(u,v,b) + (\beta \times J_F(b) \times d) & u,v \in F_{mid} \\ I(u,v,b) & u,v \notin F_{mid} \end{cases} \quad \text{Watermark bit} = 0$$

$$I_w(u,v,b) = \begin{cases} I(u,v,b) - (\beta \times J_F(b) \times d) & u,v \in F_{mid} \\ I(u,v,b) & u,v \notin F_{mid} \end{cases} \quad \text{Watermark bit} = 1$$

where $I_w(u, v, b)$ is the modified DCT coefficient in location $(u, v)$ for block $b$
$I(u, v, b)$ is the DCT coefficient in location $(u, v)$ for block $b$
$\beta$ is the scaling factor
$J_F(b)$ is the JND value generated for the block from the equation above
$d$ is the d-sequence generated
$F_{mid}$ is the middle frequencies of the DCT block

Finally, the block containing the watermarked DCT coefficients is inverse-transformed to obtain the final watermarked image. Each block containing the watermarked coefficients in the transformed domain is converted back to the image block in the pixel domain. Hence we obtain the final watermarked image. The algorithm used in MATLAB to compute the inverse DCT is shown below:

$$A_{mn} = \sum_{p=0}^{M-1} \sum_{q=0}^{N-1} \alpha_p \alpha_q B_{pq} \cos\frac{\pi(2m+1)p}{2M} \cos\frac{\pi(2n+1)q}{2N}, \quad \begin{array}{l} 0 \leq m \leq M-1 \\ 0 \leq n \leq N-1 \end{array}$$

$$\alpha_p = \begin{cases} 1/\sqrt{M} & p = 0 \\ \sqrt{2/M} & 1 \leq p \leq M-1 \end{cases} \qquad \alpha_q = \begin{cases} 1/\sqrt{N} & q = 0 \\ \sqrt{2/N} & 1 \leq q \leq N-1 \end{cases}$$

### 3.3 Watermark Detection

In order to recover the watermark we use the correlation-based watermark detection scheme. Here, the image is first broken down into the same $8 \times 8$ blocks as done in watermark embedding and then the DCT is performed on each block. The DCT coefficients of the mid-frequency values thus obtained are compared with the d-sequence that is generated using the same prime number used in watermark embedding.

$$\text{Correlation } C(b) = \frac{1}{N}(I^*(b).W(b))$$

$$\text{Recovered watermark bit} = \begin{cases} 0 & \text{if } C(b) > T \\ 1 & \text{if } C(b) < T \end{cases}$$

where, $T$ is the Threshold level
    $C(b)$ is the correlation value for block $b$
    $I^*(b)$ is the DCT coefficient of the watermarked image assumed to have been transformed by processing or attack
    $W(b)$ is the d-sequence that is generated using the same prime number that was used to generate the d-sequence at the embedding stage.

Selecting a threshold $T$ in the process of correlation filters out the unwanted noise. We must be aware that determining the presence of a watermark through correlation is a statistical test and hence there is a possibility of obtaining detection errors. Errors can be of two types, '0' that is falsely detected as '1' and '1' that is falsely detected as '0'. The setting of the threshold $T$ is considered as a decision based on the need to minimize errors, such as those mentioned above, in watermark detection. The results of our watermarking scheme are shown in the next chapter. Some watermarking attacks are conducted to test the performance of the proposed watermarking scheme.

## 4 Watermark Evaluation

Signal to noise ratio (SNR) effectively measures the quality of the watermarked image as compared to the original image. This difference is represented as an error function that shows how close the watermarked image is to the original image and it is written as:

$$e(x,y) = I(x,y) - I_W(x,y)$$

The larger the value of e (x, y) the greater is the distortion caused by the watermark and the attacks. One of the simplest distortion measures is the mean square error (MSE) function [20]:

$$MSE = \frac{\sum [f(i,j) - F(i,j)]^2}{N^2}$$

The peak signal to noise ratio (PSNR) is calculated by using the formula:

$$PSNR = 20\log_{10}\left(\frac{255}{\sqrt{MSE}}\right)$$

√MSE is called the root mean square error and 255 is the maximum value of luminance level. It should be noted that PSNR does not take aspects of the HVS into consideration although it provides an overall evaluation of the difference between the original and the watermarked image. For this reason, we will use another perceptual quality measure called the weighted peak signal to noise ratio (WPSNR). This metric takes into account the objective measure as well as the HVS. The human eye is less sensitive to changes in highly textured areas and hence an additional parameter called the noise visibility function (NVF) is introduced. This helps us calculate the change in the perceptual quality more accurately. The formula for WPSNR is shown below:

$$WPSNR = 20\log_{10}\left(\frac{255}{\sqrt{MSE} \times NVF}\right)$$

NVF uses a Gaussian model to estimate the amount of texture content in any part of an image [20]. In regions with edges and texture NVF will have a value greater than 0 while in smooth regions the value of NVF will be greater than 1. The formula to calculate this factor as a simplified function is:

$$NVF = NORM\left\{\frac{1}{1+\delta_{block}^2}\right\}$$

where δ is the luminance variance for the $8 \times 8$ block and NORM is a normalization function.

## 5 Experimental Results

Our experiments on the proposed content based watermarking are based on grayscale images. The cover objects used are the images of Lena and Boat. The prime number $q$ that is used to generate the d-sequence is 2467 and on using the appropriate scaling factor and threshold we notice that the watermark is recovered perfectly well. As mentioned earlier the scaling factor is set according to the content and the quality of the watermarked image. The greater the scaling factor, the better is the watermark detection however reducing the overall quality of the image. Hence an optimum value is chosen accordingly and it is set to 0.007 for Lena. Here, a watermark of size $12 \times 12$ pixels has been used and the peak signal to noise ratio (WPSNR) is found to be 38.99 dB.

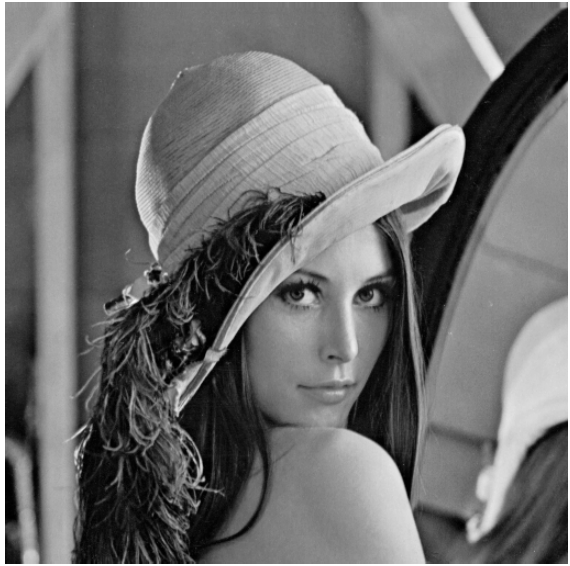

Figure 6 Lena Reference Image (512 x 512 Pixels)

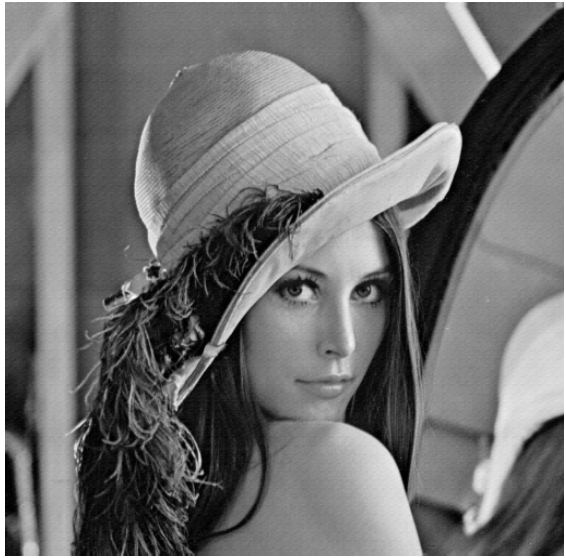

Figure 7 Watermarked Image WPSNR = 38.99 dB

| Figure 8 Original Watermark | Figure 9 Recovered Watermark |

On changing the prime number *q* to 8069 and using the same scaling factor as above we notice that the original watermark has been recovered perfectly. When we use a slightly bigger watermark of size 15×12 pixels and peak signal to noise value (WPSNR) is again found to be 38.99 dB. Note that the scaling factor is kept the same, as increasing it would result in some visible distortions in the watermarked image. The WPSNR will be the same for a given scaling factor; a change in the watermark size will not affect this value because all the DCT blocks are modified irrespective of the size of watermark, providing robustness and easy watermark detection.

The watermark is added in such a way so as to keep the amplitude of the signal below the noise distortion sensitivity of each pixel and varies very slightly with the size of the watermark for a given image. The detection statistic does not change much for the two images because there is minimal change in the size of the watermark.

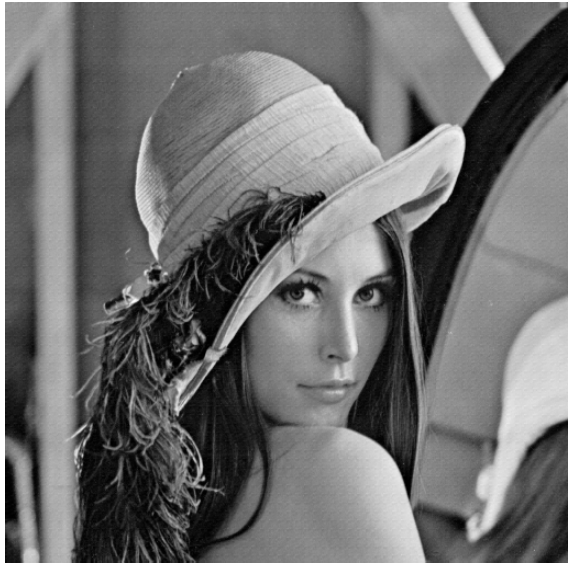

Figure 10 Watermarked Image WPSNR =38.79 dB

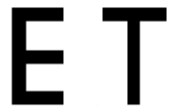
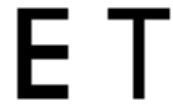

Figure 11 Embedded Watermark　　　　Figure 12 Recovered Watermark

Now, we use a different cover image with the same two prime numbers that was used for the earlier result; the cover image is Boat (Figure 13), and the watermark is of size $64 \times 64$. We observe that there is a change in the value of WPSNR for Boat as this image has a greater scaling factor. This value is decided by observing the quality of the watermarked image. The scaling factor for every image is dependent on the image content and varies according to the properties that are used to determine the mask namely, texture, corner, edge and luminance information.

For this image the scaling factor suggested is 0.084. The watermark is recovered, although some amount of noise is present in the recovered watermark and it is seen that this scheme holds good for different sizes of watermark. As observed, the scaling factor that is used to watermark the cover object allows enough information to be added without any visible distortion in the final watermarked image while also allowing for a good recovery of the watermark.

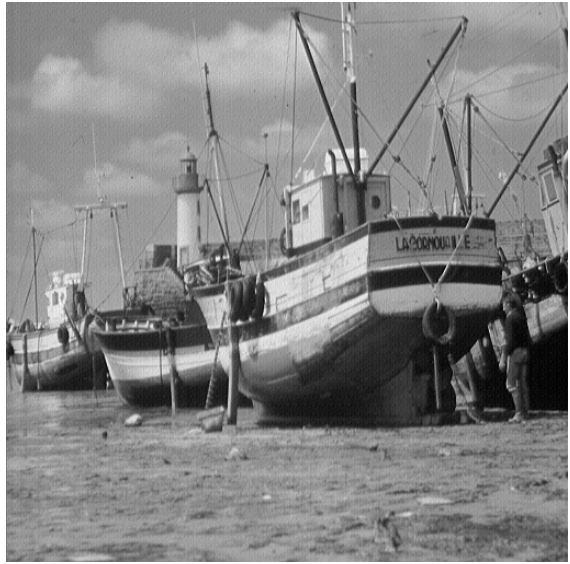

Figure 13 Watermarked Image WPSNR = 35.54

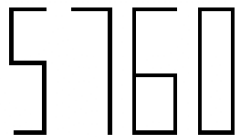

Figure 14 Embedded Watermark

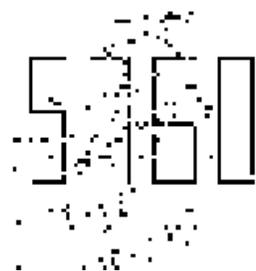

Figure 15 Recovered Watermark

We experimented with various combinations of primes to verify the robustness of our scheme and found that in almost all the combinations the detection of the watermarks was highly satisfactory. The WPSNR value for the relatively smaller watermarks has been found to be same and it slightly decreases for larger watermarks. Similarly the WPSNR value remains constant for all the prime numbers. The correlation threshold is entirely dependent on the scaling factor, which in turn is dependent on the image content. This factor is fixed for an image and value is fixed based on the desire to minimize false alarms and false rejections.

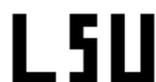

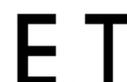

Figure 16 Watermark ($12 \times 12$)

Figure 17 Watermark ($15 \times 12$)

Figure 18 Watermark ($32 \times 32$)       Figure 19 Watermark ($64 \times 64$)

     The change in WPSNR values with varying scaling factors for images Lena and Boat may be observed from Figure 20. From this plot we see that the rate at which the PSNR decreases for increasing values of the scaling factor is unchanged but the huge gap between the values suggest the difference in the JND values due to the image content. The WPSNR for the scaling factor used for Boat is 35.54, but in the case of watermarking Lena, for the same value the image is highly distorted.

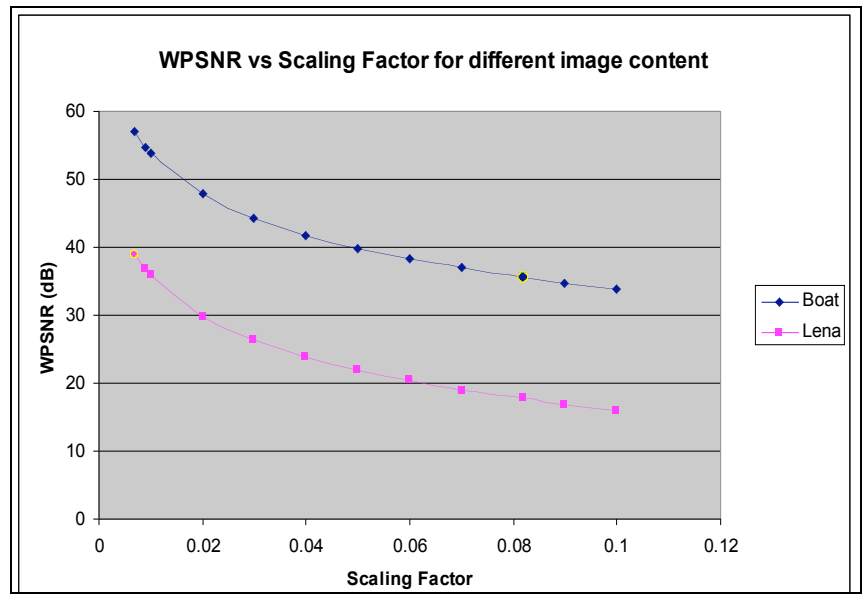

Figure 20 WPSNR vs. Scaling Factor Plot for Lena and Boat

     The JND mask value for two randomly selected rows for Lena is shown in Figure 21. This plot signifies the maximum weight of the watermark allowed to be added to the image without a visible distortion. A similar plot for the same two rows for the Boat image is shown in Figure 22. We can notice the change in this plot for these two images. It is also seen that the values are equally high for both Lena and Boat in row 32 and hence more watermark information can be embedded in these areas.

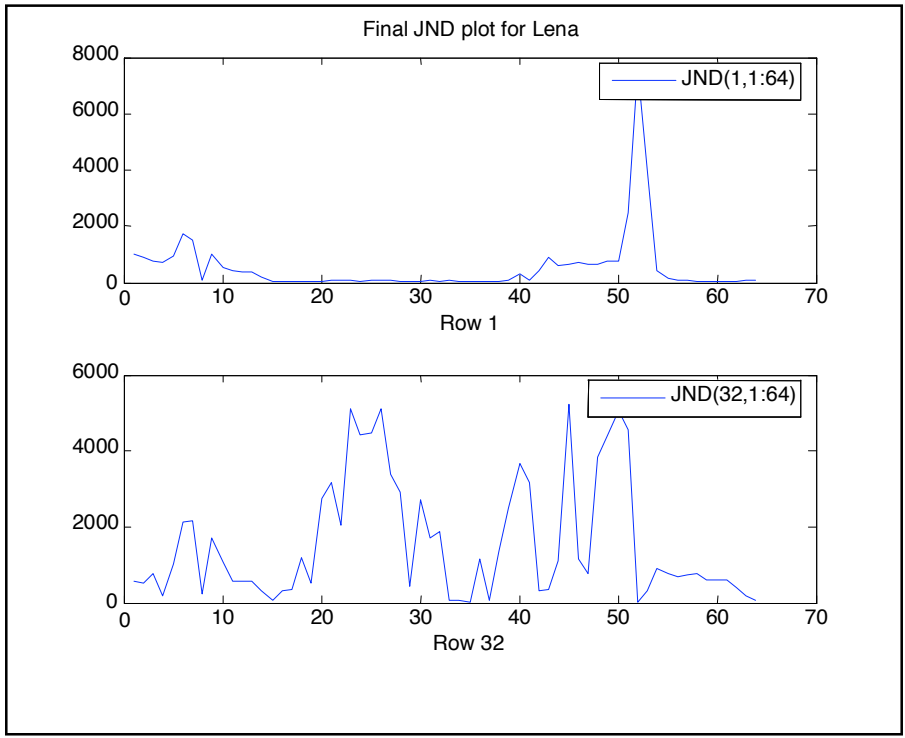

Figure 21 Normalized JND values for Lena

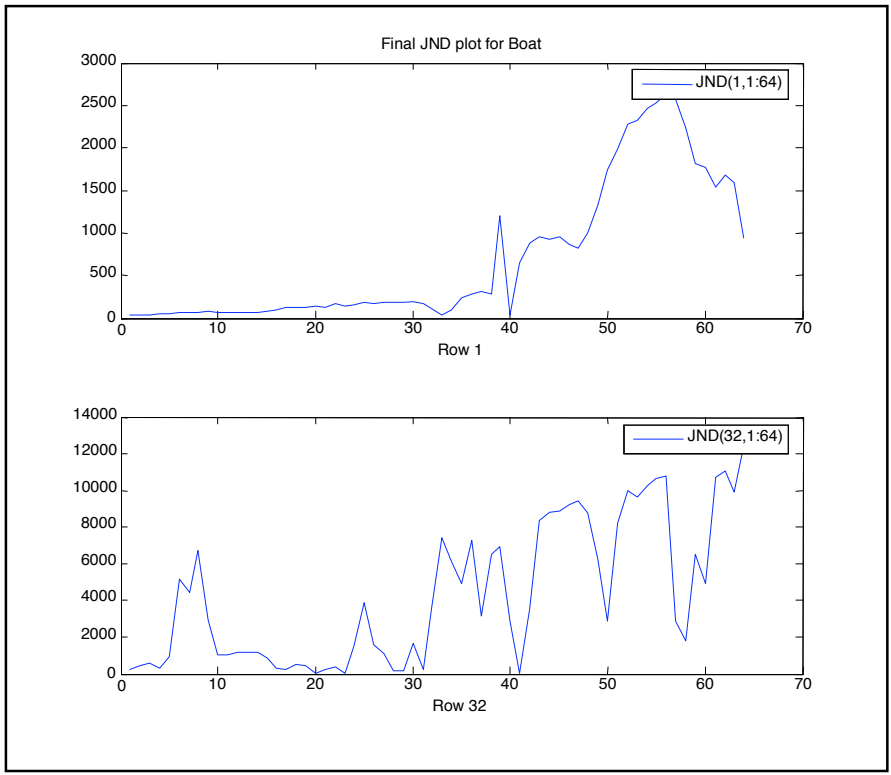

Figure 22 Normalized JND values for Boat

Since, the watermark is spread throughout the cover object based on the frequency; an attack that is aimed at the watermark removal cannot be successful unless it attacks the fundamental structure of the image itself, which would result in a highly degraded image. Since the watermarking is performed only to the mid-frequency components of the image blocks, the effect of most compression algorithms that usually target the high frequency components is largely avoided and hence the high robustness against compression.

## 6 Attacks and Analysis of Results

Various attacks are performed to test the robustness of our scheme and it is found, that our scheme performs excellently against JPEG compression and moderately well against common spatial attacks. The block size was kept constant at $8 \times 8$, mostly in anticipation of the JPEG compression. The payload as well as the robustness can be increased by also embedding in either the low or the high frequency components of the DCT block, depending on the attack. We have made selective use of Stirmark benchmarking technique [21], [22], to test the robustness of our scheme for JPEG compression and median filter attack. The other attacks that are performed on the watermarked images are introduction of Gaussian and salt and pepper noise and use of image filter. The result of the JPEG compression for different quality factors *(q)* ranging from 45 to 30 is shown below along with the recovered watermark. It should be noted that this scheme performs excellently for JPEG compression of quality factor 45 and above.

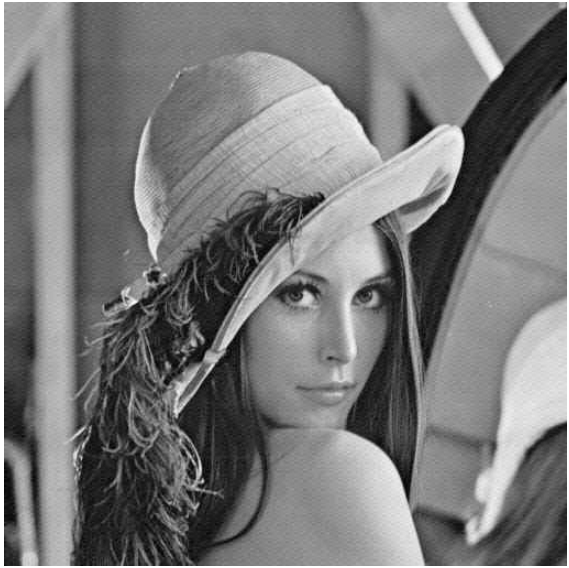
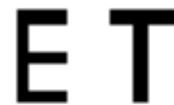

Figure 23 JPEG Compression (*q* = 45)          Figure 24 Recovered Watermark

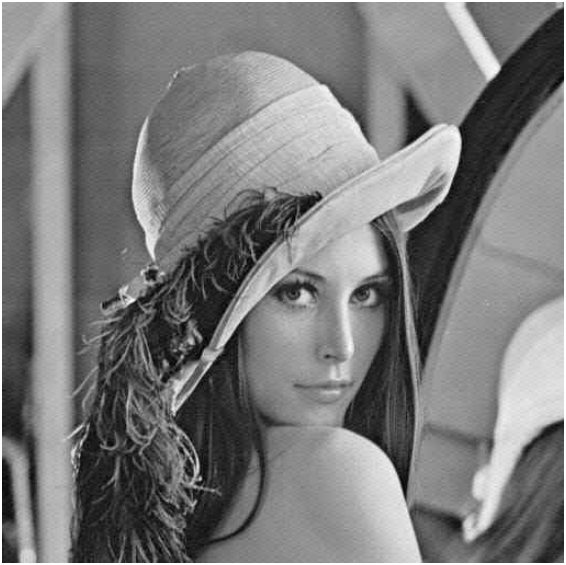
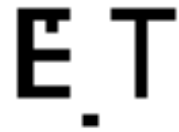

Figure 25 JPEG Compression (*q*= 40)         Figure 26 Recovered Watermark

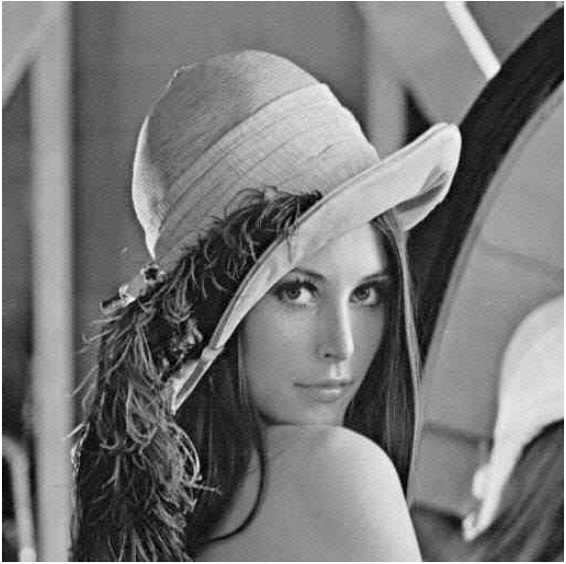
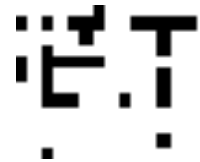

Figure 27 JPEG Compression (q= 35)         Figure 28 Recovered Watermark

    One of the most interesting results is that the recovered watermark after JPEG compression with a quality factor of 45 is sometimes much better than the watermark that is recovered from a pristine watermarked image. The reasoning behind this is that, the detection errors in the unmodified source are deemed to be right on the correlation boundary and the addition of noise is just enough to push them over the edge [23].

From the above results we can say that for JPEG compression with a quality factor of 40, the watermark detection and extraction is near perfect. The recovered watermark for a quality factor of 35 shows a number of detection errors and this only becomes highly noticeable for a quality factor of 30. The overall robustness of our scheme for JPEG compression is considered high level, according to the robustness requirements table provided by Petitcolas [22].

We then test our scheme for its robustness against different types of noise. This is done by first introducing noise into the watermarked Lena of size $512 \times 512$. Gaussian noise with zero mean is introduced to verify as to what extent our proposed scheme can withstand noise. From the results shown below, we can observe that for a Gaussian noise of 2 %, the watermark recovery is moderate, with very few detection errors. We must keep in mind that most DCT block based schemes offer moderate robustness to noise and less robustness to common geometric attacks like scaling. An alternate technique would be to employ dual watermarking, in both frequency as well as spatial domain using CDMA spread spectrum. But this would not serve the purpose of imperceptibility of the watermark and as it is widely known; CDMA spread spectrum techniques result in visible distortion, have limited capacity and high processing requirements [23]. This is due to the exhaustive search performed to detect the embedded sequence over each pixel of the image.

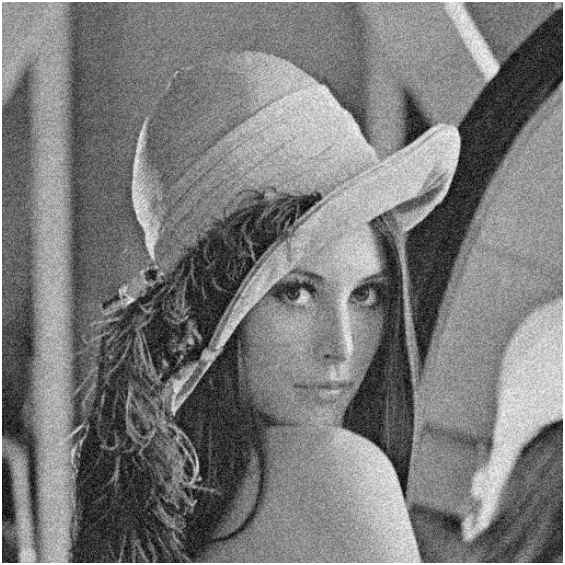 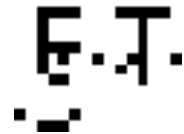

Figure 29 Uniform Gaussian Noise 2%        Figure 30 Recovered Watermark

Another common attack on images is filtering. We test our watermarking scheme for its robustness to median filtering attack generated using Stirmark [21], [22]. Median filter is similar to an averaging filter; each pixel output is set to the median of the pixel values in the neighborhood of the corresponding input pixel, as specified by the window size. The window size of $3 \times 3$ is used for our experiments. This is considered as moderate robustness for any watermarking scheme. As we can see, the watermark has been recovered almost perfectly except for some detection errors, which are introduced due to the filter.

## 7 Conclusions

This paper provides a comprehensive evaluation and implementation of a content based watermarking scheme that improves upon the earlier work of Kay and Izquierdo. By analyzing the cover object in both frequency and spatial domains, a distortion sensitivity of the image content is determined. Local information that is derived from properties such as texture, corner, edge and luminance is used to determine a mask of just noticeable difference values. This value, which is based on the image content, determines the strength of the watermark information that will be embedded. Our observations regarding the proposed watermarking scheme are summarized below:

- We employ a better method of detecting edges using phase congruency, allowing us to detect more edges accurately.
- Our scheme implements an algorithm that detects corners using curvature scale space instead of a Moravec operator. The detected corner is used as a factor to establish the uniform regions in the image, which is utilized to determine the JND mask.
- The robustness of our scheme to JPEG compression is found to be very good at a quality factor of 40 and reasonably good at a quality factor of 35. The results for other image processing attacks like median filtering and contrast-sharpening filter were also found to be good, although it is not very robust against scaling and high noise levels.
- Our scheme introduces a content based watermarking scheme using decimal sequences and the results are found to be highly satisfactory in terms of watermark detection. Any random sequence may be used to embed the watermark and the decision of using decimal sequences is based on the ease with which it can be generated, requiring only a prime number. Also, more flexibility can be achieved with the choice of various prime numbers that can be used for this purpose.
- A very good balance between robustness and imperceptibility has been achieved using this scheme as observers can evaluate the quality of the watermarked image as well as the recovered watermark to be good. Experimentation using various sizes of watermarks and different images enables a better understanding of the scheme. WPSNR is used to evaluate the perceptual quality of the watermarked image effectively and accurately, considering the effect of HVS.

Although this paper was limited to watermarking of gray scale images in the DCT domain, further research can be done in implementing content based watermarking using decimal sequence for color images and video watermarking. To increase the scheme's robustness against geometrical attacks like scaling, cropping as well as higher noise levels, we suggest an implementation of a hybrid watermarking scheme in both the spatial as well as the frequency domain, which need not be restricted to DCT. CDMA spread spectrum approach to the hybrid scheme may also be considered although this may not quite serve the purpose of imperceptibility of the watermark.